\documentclass[11pt,a4paper]{article}
 \pdfoutput=1

\usepackage{jheppub_kim}
\usepackage{rotating} 
\usepackage{graphicx,epsfig}
\usepackage{amsmath}
\usepackage {amssymb}
\usepackage{subfigure}

\usepackage{relsize}

\providecommand{\U}[1]{\protect\rule{.1in}{.1in}}

\newcommand{\newc}{\newcommand}

\newc{\be}{\begin{equation}}
\newc{\ee}{\end{equation}}

\newc{\ba}{\begin{eqnarray}}
\newc{\ea}{\end{eqnarray}}
\newc{\bea}{\begin{eqnarray*}}
\newc{\eea}{\end{eqnarray*}}
\newc{\D}{\partial}
\newc{\ie}{{\it i.e.} }
\newc{\eg}{{\it e.g.} }
\newc{\etc}{{\it etc.} }
\newc{\etal}{{\it et al.}}
\newc{\lcdm}{$\Lambda$CDM }

\newc{\ra}{\Rightarrow}

\title{Observational constraints on Barrow holographic dark energy}

\author[a]{Fotios K. Anagnostopoulos}

\author[b]{Spyros Basilakos}

\author[c,d,e]{Emmanuel N. Saridakis}

\affiliation[a]{Department of Physics, National \& Kapodistrian University of 
Athens, 
Zografou Campus GR 157 73, Athens, Greece}

\affiliation[b]{Academy of Athens, Research Center for Astronomy and
Applied Mathematics, Soranou Efesiou 4, 11527, Athens, Greece}

\affiliation[c]{National Observatory of Athens, Lofos Nymfon, 11852 Athens, 
Greece}

 \affiliation[d]{Department of Physics, National Technical University of 
Athens, 
Zografou
Campus GR 157 73, Athens, Greece}
\affiliation[e]{Department of Astronomy, School of Physical Sciences, 
University of Science and Technology of China, Hefei 230026, P.R. China}

\emailAdd{svasil@academyofathens.gr}
\emailAdd{msaridak@phys.uoa.gr}

\abstract{ 
 We use observational data from Supernovae (SNIa) Pantheon sample, 
as well as from direct measurements of the Hubble parameter from the 
cosmic chronometers (CC) sample, in order to extract constraints on the 
scenario 
of Barrow holographic dark energy.  The latter is a holographic dark energy 
model based on the recently proposed Barrow entropy, which arises from the 
modification of the black-hole surface due to quantum-gravitational effects.  
We first consider the case where the new deformation exponent $\Delta$ is the 
sole model parameter, and we show that although the standard value $\Delta=0$, 
which corresponds to zero deformation, lies 
within the 1$\sigma$ region, a deviation is favored. In the case where we let 
both $\Delta$
and the second model parameter to be free we find that a deviation from  
standard holographic dark energy is preferred.  Additionally, applying 
the Akaike, Bayesian
and Deviance 
Information Criteria, we conclude that the one-parameter model  is statistically 
compatible with 
$\Lambda$CDM paradigm, and preferred
comparing to the two-parameter one. Finally, concerning the present 
value of the Hubble parameter we find that it is close to the Planck  
value. }


\begin{document}
\maketitle

\section{Introduction}

Accumulated data from various probes lead to the safe deduction that the 
universe have undergone two phases of accelerated expansion, at early and late 
cosmological times respectively. Such a behavior may require the 
introduction of extra degrees of freedom that are capable of triggering it (the 
simple cosmological constant can sufficiently describe the latter phase, but it 
is not adequate to describe the former one). A first main direction is the 
construct modified gravitational theories, that posses general relativity as a 
particular limit, but which on larger scales can produce the above 
phenomenology, such as in   $f(R)$  gravity
\cite{Starobinsky:1980te,DeFelice:2010aj,Nojiri:2010wj},
  $f(G)$ gravity \cite{Nojiri:2005jg},  Galileon theory 
\cite{Deffayet:2009wt},  
$f(T)$ gravity  
\cite{Ben09,Chen:2010va,Kofinas:2014owa}, Finsler gravity 
\cite{Basilakos:2013hua} etc (see 
\cite{Nojiri:2006ri,Capozziello:2011et,Cai:2015emx} for reviews). 
The second main direction is to maintain general relativity as the underlying 
gravitational theory and introduce the  the 
inflaton  field(s) \cite{Olive:1989nu,Bartolo:2004if} and/or the dark 
energy  concept attributed to new fields, 
particles or fluids \cite{Copeland:2006wr,Cai:2009zp}.

One interesting approach for the description of dark energy arises from 
holographic considerations 
\cite{tHooft:1993dmi,Susskind:1994vu,Bousso:2002ju,Fischler:1998st,
Horava:2000tb}. Specifically, since the largest length of a quantum field 
theory is connected to its Ultraviolet cutoff \cite{Cohen:1998zx}, one can 
result to a vacuum energy which at cosmological scales   
forms a form of holographic dark energy \cite{Li:2004rb,Wang:2016och}.
Holographic dark energy  is very efficient in quantitatively describe the 
late-time acceleration 
\cite{Li:2004rb,Wang:2016och,Horvat:2004vn,Huang:2004ai,Pavon:2005yx,
Wang:2005jx,
Nojiri:2005pu,Kim:2005at,
Wang:2005ph, Setare:2006wh,Setare:2008pc,Setare:2008hm} and it is in 
agreement  with   
observational data 
\cite{Zhang:2005hs,Li:2009bn,Feng:2007wn,Zhang:2009un,Lu:2009iv,
Micheletti:2009jy,DAgostino:2019wko,Sadri:2019qxt,Molavi:2019mlh}. Hence, many 
extensions of the basic scenario have appeared in the literature, based mainly 
on the use of different horizons as the largest distance (i.e. the universe 
``radius'')  
\cite{Gong:2004fq,Saridakis:2007cy,  
Setare:2007we,Cai:2007us,Setare:2008bb,Saridakis:2007ns,Saridakis:2007wx,
Jamil:2009sq,
Gong:2009dc, 
Suwa:2009gm,BouhmadiLopez:2011xi,Malekjani:2012bw,
Khurshudyan:2014axa,
Landim:2015hqa,Pasqua:2015bfz,
Jawad:2016tne,Pourhassan:2017cba,Nojiri:2017opc,Saridakis:2017rdo,
Saridakis:2018unr,Aditya:2019bbk,Nojiri:2019kkp,Geng:2019shx,Waheed:2020cxw}.

One such extension is Barrow holographic dark energy, which arises by applying 
the usual holographic principle but using the recently proposed Barrow entropy 
instead of the Bekenstein-Hawking one.
The later is a modification of the black-hole entropy caused by 
quantum-gravitational effects that deform the horizon, leading it to acquire  a 
fractal, intricate,  structure \cite{Barrow:2020tzx}. Hence, one results with 
an extended holographic dark energy, which includes basic holographic dark 
energy as a sub-case in the limit where Barrow entropy becomes  the 
Bekenstein-Hawking one, but which in general is a novel scenario which  
exhibits more interesting and richer 
phenomenology \cite{Saridakis:2020zol}.

In the present work  we desire to use observational data from from Supernovae 
(SNIa) Pantheon sample, and from direct Hubble 
constant measurements with cosmic chronometers (CC), in order to 
constrain Barrow holographic dark energy, and in particular to impose 
observational bounds in the new Barrow exponent that quantifies the 
quantum-gravitational deformation and thus the deviation from usual holographic 
dark energy. The plan of the work is the following: In Section \ref{model} we 
briefly review Barrow holographic dark energy. In Section \ref{Methods} we 
present the various datasets, the applied 
methodology, and the information criteria that we will use. In Section  
\ref{Observcion} we provide the obtained results and we give the corresponding 
contour plots. Finally, in Section \ref{Conclusions} we summarize and conclude.

\section{Barrow holographic dark energy}
\label{model}

In this section we present the cosmological scenario of Barrow holographic dark 
energy. Barrow entropy is a quantum-gravitationally corrected black-hole 
entropy due to the fractal structure brought about in its horizon, and it takes 
the form \cite{Barrow:2020tzx}
\begin{equation}
\label{Barrowentropy}
S_B=  \left (\frac{A}{A_0} \right )^{1+\Delta/2}, 
\end{equation}
where $A$ is the standard horizon area  and $A_0$ the Planck area. The quantum 
deformation, and hence the deviation from Bekenstein-Hawking entropy is 
quantified by the new exponent $\Delta$, which takes the value  $\Delta=0$ in 
the standard, non-deformed case, while  for  $\Delta=1$ it corresponds to   
maximal deformation.
 
We consider a flat   Friedmann-Robertson-Walker (FRW) 
  geometry  with metric
\begin{equation}
\label{FRWmetric}
ds^{2}=-dt^{2}+a^{2}(t)\delta_{ij}dx^{i}dx^{j}\,,
\end{equation}
where $a(t)$ is the scale factor. As it was shown in \cite{Saridakis:2020zol}, 
application of the holographic principle but using   Barrow entropy 
(\ref{Barrowentropy}), leads to Barrow holographic dark energy, whose energy 
density reads: 
 \begin{equation}
\label{FRWBarrowHDE}
\rho_{DE}={C} R_h^{\Delta-2}.
\end{equation}
where  ${C}$ is a parameter with dimensions  $[L]^{-2-\Delta}$, and $R_h$ 
 the   future event horizon 
\begin{equation}
\label{eventfuturehor}
R_h\equiv a\int_t^\infty \frac{dt}{a}= a\int_a^\infty \frac{da}{Ha^2},
\end{equation}
 where $H\equiv \dot{a}/a$ is the Hubble parameter.

The two  Friedmann equations are  
 \begin{eqnarray}
\label{Fr1bFRW}
3M_p^2 H^2& =& \ \rho_m + \rho_{DE}    \\
\label{Fr2bFRW}
-2 M_p^2\dot{H}& =& \rho_m +p_m+\rho_{DE}+p_{DE},
\end{eqnarray}
with $M_p=1/\sqrt{8\pi G}$ the Planck mass.
Moreover, $p_{DE}$ is the pressure of  Barrow holographic dark energy, and 
$\rho_m$, $p_m$ are respectively 
 the energy density and pressure of the  matter fluid. As usual we consider the 
two sector to be non-interacting, and thus  the usual
conservation 
equations hold
\begin{eqnarray}\label{rhoconservFRW}
&&\dot{\rho}_m+3H(\rho_m+p_m)=0,\\
&&\dot{\rho}_{DE}+3H\rho_{DE}(1+w_{DE})=0.
\label{DErhoconservFRW}
\end{eqnarray}
In the following we focus on the case of dust matter, namely we assume that 
$p_m=0$.

Introducing the density parameters  
$ \Omega_i\equiv\frac{1}{3M_p^2H^2}\rho_i$, in the case $0\leq\Delta<1$ one can 
easily extract the evolution equation for $\Omega_{DE}$ as a function of 
$x\equiv\ln a=-\ln(1+z)$, with $z$ the redshift (with $a_0=1$), namely 
\cite{Saridakis:2020zol}
  \begin{eqnarray}\label{Odediffeq}
&&
\!\!\!\!\!\!\!\!\!\!\!\!\!
\frac{\Omega_{DE}'}{\Omega_{DE}(1-\Omega_{DE})}=\Delta+1+
Q
(1-\Omega_{DE})^{\frac{\Delta}{
2(\Delta-2) } } 
 (\Omega_{DE})^{\frac{1}{2-\Delta } } 
e^{\frac{3\Delta}{2(\Delta-2)}x},
\end{eqnarray}
 with
   \begin{equation}\label{Qdef}
Q\equiv (2-\Delta)\left(\frac{{C}}{3M_p^2}\right)^{\frac{1}{\Delta-2}} 
\left(H_0\sqrt{\Omega_{m0}}\right)^{\frac{\Delta}{2-\Delta}}
\end{equation}
a dimensionless parameter and   where primes denote derivatives with respect to 
$x$.  Furthermore, 
    the  equation of state 
  for Barrow holographic dark energy, i.e $w_{DE}\equiv p_{DE}/\rho_{DE}$, is 
given by   
\begin{equation}\label{wDEFRW}
w_{DE}=-\frac{1\!+\!\Delta}{3}
-\frac{Q}{3}
(\Omega_{DE})^{\frac{1}{2-\Delta } } (1\!-\!\Omega_{DE})^{\frac{\Delta}{
2(\Delta-2) } }
e^{\frac{3\Delta}{2(2-\Delta)}x}.
\end{equation}


Barrow holographic dark energy is a new dark energy scenario. In the case 
$\Delta=0$ it coincides
 with  standard holographic dark energy  $\rho_{DE}=3c^2 M_p^2 R_h^{-2}$, 
with ${C}=3  c^2  M_p^2$   the model parameter. In this case (\ref{Odediffeq})
becomes  
$\Omega_{DE}'|_{_{\Delta=0}}= 
\Omega_{DE}(1-\Omega_{DE})\left(1+2\sqrt{\frac{3M_p^2\Omega_{DE}}{{C}}}
\right)
$, and can be analytically solved  implicitly
\cite{Li:2004rb}, while  
$w_{DE}|_{_{\Delta=0}}=-\frac{1}{3}-\frac{2}{3}\sqrt{\frac{3M_p^2 
\Omega_{DE}}{{C}}}$, which is the standard holographic 
dark energy result
\cite{Wang:2016och}.
 However, in the case $\Delta>0$, where the deformation effects 
 switch on, the scenario at hand departs 
from the standard one, leading to different cosmological behavior. Lastly, in 
the upper limit  $\Delta=1$, it coincides with $\Lambda$CDM cosmology.

\section{Data and Methodology}
\label{Methods}

 In this section we provide the various data sets that are going to be used for 
the observational analysis, and then   we present the statistical methods that 
we   employ. 
We   use     data from Supernovae type 
Ia  observations together with direct $H(z)$ Hubble data, and we   apply the 
method of  maximum likelihood analysis to 
 in order to extract constraints on the free model parameters. As a final step, 
we will employ known information criteria in order to
    assess the quality of the fittings.

\subsection{Cosmological probes}

\subsubsection{Type Ia Supernovae}
Perhaps the most known and frequently used cosmological probe are distant Type 
Ia Supernovae. A supernova explosion is an extremely luminous event, with its 
brightness being comparable with the brightness of its host galaxy
\cite{Scolnic:2017caz}. The observed light curves posses 
peak brightness mostly unaffected by the distance, thus can be used as standard 
candles. Specifically, one could use the observed distance modulo, $\mu_{obs}$, 
to constrain cosmological models. We use the most recent data set available, 
namely the binned Pantheon dataset  described at \cite{Scolnic:2017caz}. 
Finally, the corresponding likelihood reads
\begin{equation}
    \mathcal{L}_{SNia}(Y;\mathcal{M})\sim \exp\left(-\frac{1}{2}\sum_{i=1}^{40} 
m_{i}C_{cov}^{-1}m_{i}^{\dagger}\right),
\end{equation}
where Y is the vector of the free parameters of the cosmological model, $m_{i} 
=\mu_{obs,i}-\mu_{theor}(z_i)-\mathcal{M}$ and $\mu_{theor} = 
5\log(\frac{D_{L}}{1Mpc}) + 25$, and $D_L$ is the standard luminocity distance, 
given 
as $D_L = c(1+z)\int_{0}^{z}\frac{1}{H(z)}$, that holds for a flat FRWL 
space-time, regardless of the underlying cosmology. Finally, $C_{cov}$ is the 
covariance matrix of the binned Pantheon dataset. The parameter $\mathcal{M}$ 
is 
an intrinsic free parameter to the Pantheon dataset and quantifies a variety of 
observational uncertainties, i.e host galaxy properties, etc. 
\subsubsection{Cosmic chronometers}
Data from the so-called ``cosmic chronometers'' (CC), are measurements of the 
Hubble rate, based upon the estimation of the \emph{differential age} of 
passive 
evolving galaxies. The latter are galaxies with their emission spectra    
dominated by   old stars population. The central idea is to use the definition 
of the Hubble rate, re-parametrized in terms of redshift, i.e
\begin{equation}
    H \equiv \frac{\dot{a}}{a} = - \frac{1}{1+z} \frac{dz}{dt}.
\end{equation}
From this point, the redshift is relatively easily observed spectroscopically 
and the remaining work is to estimate the quantity $dz/dt$. As it was firstly 
proposed by Jimenez and Loeb in \cite{Jimenez:2001gg}, this is possible via 
measuring the age difference between two sets of passively evolving galaxies, 
lying within a small redshift difference. The observational method and specific 
information from an astrophysical point of view are described in detail in 
\cite{Moresco:2018xdr,Moresco:2020fbm}. 

From a cosmological viewpoint, it is important to note that data from cosmic 
chronometers are essentially model independent, as long as we work within an 
FRWL space-time without extrinsic curvature. Furthermore, the redshift range of 
the available cosmic chronometers extends to 2, thus they allow for more 
stringent 
constraints to the cosmological models under study. Thus, cosmic chronometers 
are used widely in the field 
\cite{Haridasu:2018gqm,Anagnostopoulos:2018jdq,DAgostino:2019wko,Ryan:2019uor}. 
In this work the sub-sample of \cite{Farooq:2016zwm}, consisting of only CC 
data, 
is employed. The likelihood for the cosmic chronometers, assuming gaussian 
errors, reads 
\begin{equation}
    \mathcal{L}_{CC}(Y) \sim \exp\left[ -\frac{1}{2}\sum_{i=0}^{31} 
\frac{\left(H(z_i)_{theor} - H_{obs,i}\right)^2}{\sigma_{i}^2}\right],
\end{equation}
where $\sigma_{i}$ are the corresponding 
errors.

\subsubsection{Joint analysis}

In order to obtain the   joint observational constraints on the cosmological 
scenario by using     $P$ cosmological 
datasets, we first  introduce the total likelihood function as  
 \begin{equation}
  \mathcal{L}_{\text{tot}}(Y) = \prod_{p=1}^{P} 
\mathcal{L}_{i},
  \end{equation}
   assuming Gaussian errors, and where no correlation between various data sets 
employed.
 Hence, the total  $ \chi_{\text{tot}}^2$ function will be
        \begin{equation}
        \chi_{\text{tot}}^2 = \sum_{p=1}^{P}\chi^2_{P}\,.
        \end{equation}
 The parameter vector has dimension $k$, namely the $\nu$
 parameters of the scenario, plus the number of 
hyper-parameters  $\nu_{\text{hyp}}$ of the applied datasets, i.e.
$k = \nu + 
\nu_{\text{hyp}}$.
For the scenario of Barrow  holographic dark energy, and since we are using   
Hublle rate and  SNIa 
data, the free parameters are contained in the vector 
$a_m = (\Omega_{m0},C,\Delta, h,\mathcal{M})$, with  
$h=H_{0}/100$. We apply the Markov Chain
Monte Carlo (MCMC) algorithm in the environment of the Python package emcee
\cite{ForemanMackey:2012ig}, and we perform the minimization of $\chi^2$ with 
respect to $a_m$.    We use 800 chains (walkers) and 
3500 steps (states).
Lastly, the convergence of the algorithm  is verified using auto-correlation 
time considerations, and additionally we 
  employ the Gelman-Rubin 
criterion  \cite{Gelman:1992zz} too for completeness.

\subsection{Information Criteria and Model Selection}
 
As a final step, we apply the known Akaike Information Criterion (AIC)
\cite{Akaike1974} and  the Bayesian Information Criterion (BIC) 
\cite{Schwarz1978},
and the Deviance 
Information Criterion \cite{Spiegelhalter2002},
in order to examine the quality 
     of the fittings and hence the relevant observational  compatibility of the 
scenarios.

The AIC is based on information theory, and it is an estimator of the 
Kullback-Leibler information with the 
property of asymptotically unbiasedness. Under the standard assumption of 
Gaussian 
errors, the corresponding estimator reads as \cite{Ann2002,Ann2002b}
\begin{equation}
 \text{AIC}=-2\ln(\mathcal{L}_{\text{max}})+2k+
 \frac{2k(k+1)}{N_{\rm tot}-k-1}\,,
 \end{equation}
with
$\mathcal{L}_{\text{max}}$   the maximum likelihood of the datasets 
  and $N_{\rm tot}$   the total  data points. 
 For large number of data points $N_{\rm tot}$ it
 reduces to $\text{AIC}\simeq -2\ln(\mathcal{L}_{\text{max}})+2k$.    On the 
other hand, the BIC criterion is an estimator of the Bayesian evidence  
\cite{Ann2002,Ann2002b,Liddle:2007fy}, given by
\begin{equation}
\text{BIC} = -2\ln(\mathcal{L}_{\text{max}})+k \,{\rm log}(N_{\text{tot}})\,.
\end{equation}

Finally, the DIC criterion is based on concepts from both Bayesian 
statistics and 
information theory \cite{Spiegelhalter2002}, and it is written as  
\cite{Liddle:2007fy}
\begin{equation}
{\rm DIC} = D(\overline{a_m}) + 2C_{B}.
\end{equation}
The variable $C_{B}$ is the  Bayesian complexity given as
$C_{B} = \overline{D(a_m)} - D(\overline{a_m})$,
with overlines denoting the standard mean value.
Moreover, $D(a_m)$ is the Bayesian Deviation, a quantity closely related to 
the 
effective 
degrees of freedom  \cite{Spiegelhalter2002}, which for the general class 
of exponential
distributions,  it reads as $D(a_m) = 
-2\ln(\mathcal{L}(a_m))$.

In order to compare a set of $n$ models we utilize the above criteria by 
extracting the relative difference of the involved IC values  $\Delta 
\text{IC}_{\text{model}}=\text{IC}_{\text{model}}-\text{IC}_{\text{min}}$,
where   $\text{IC}_{\text{min}}$ is the minimum $\text{IC}$ value in the set 
of 
compared models \cite{Anagnostopoulos:2019miu}. 
We then assign a ``probability of correctness''   to each model using the 
  rule
\cite{Ann2002,Ann2002b}
\begin{equation}
\label{prob_per_model}
P \simeq \frac{e^{-\Delta \text{IC}_{i}}}{\sum_{i=1}^{n}e^{-\Delta 
\text{IC}_{i}} },
\end{equation}
with $i$ running over the set of $n$ models. The quantity
$P$ can be considered as a measure for 
the relative strength of observational support between these two models. 
In particular, employing the Jeffreys scale 
\cite{Jeffreys,Kass:1995loi}, 
the condition $\Delta\text{IC}\leq 2$  implies statistical compatibility 
of the 
model at hand with the reference model,   the condition 
$2<\Delta\text{IC}<6$ corresponds to a middle tension between the two models, 
while  
$\Delta\text{IC}\geq 10$ implies a strong tension.

\section{Observational constraints}
\label{Observcion}

 In this section we confront the scenario of Barrow holographic dark energy 
with cosmological data from Supernovae type Ia  observations as well as from 
direct 
measurements of the Hubble rate, i.e.
$H(z)$ data, under the  procedure described above. We are interested in 
extracting the constraints on the basic model parameter $\Delta$, which 
quantifies the deviation from standard entropy, as well as on the secondary 
parameter $C$. We start by performing the analysis keeping $C$ fixed 
to the value ${C}=3$ in $M_p^2$ units, that is to the value for which  Barrow 
holographic dark energy restores exactly   standard holographic dark energy 
in the limit $\Delta=0$. In this case we can investigate purely the effect and 
the implications of the Barrow exponent $\Delta$. Additionally, as a next step 
we perform the full fitting procedure, handling both $\Delta$ and  $C$ as free 
parameters.

 \begin{table*}
\tabcolsep 4.pt
\vspace{1mm}
\resizebox{\columnwidth}{!}{
\begin{tabular}{ccccccc} \hline \hline
Models & $\Omega_{m0}$ & $h$ & $C $  & $\Delta $&  $\mathcal{M}$ 
 & $2\text{ln}\mathcal{L}_{max}$ \vspace{0.05cm}\\ 
\hline
\hline
BHDE$|_{_{C fixed}}$ & $ 0.285_{-0.044}^{+0.043}$ & $ 0.6895_{-0.0189}^{+0.0187}  $ & 
3 &   $0.095_{-0.100}^{+0.093}$ &$ -19.390_{-0.055}^{+0.056}  $ & 53.843 $ 
$ 
\\ 
   
BHDE$|_{_{C free}}$ & $0.284_{-0.044}^{+0.043}  $ & $ 
0.6892_{-0.0189}^{+0.0187} 
  $ &       $3.421_{-1.611}^{+1.753}   
$     & $ 0.094_{-0.101}^{+0.094} $       &  $ -19.390_{-0.056}^{+0.055} 
$ & 53.978  $ 
$ 
\\ 
$\Lambda CDM$ & $0.300_{-0.021}^{+0.022}$ & $0.6907_{-0.0196}^{+0.0200}$ & 
-  & -
&$-16.996_{-0.059}^{+0.057} $&54.003\\ %
\hline\hline
\end{tabular}}
\caption[]{Observational constraints on the parameters of Barrow holographic 
dark energy (BHDE), and the
corresponding $\mathcal{L}_{\text{max}}$, using SN Ia and CC datasets. }
\label{tab:Results1}
\end{table*}
 
 \begin{figure}[ht]
\centering\includegraphics[width=0.9\textwidth]{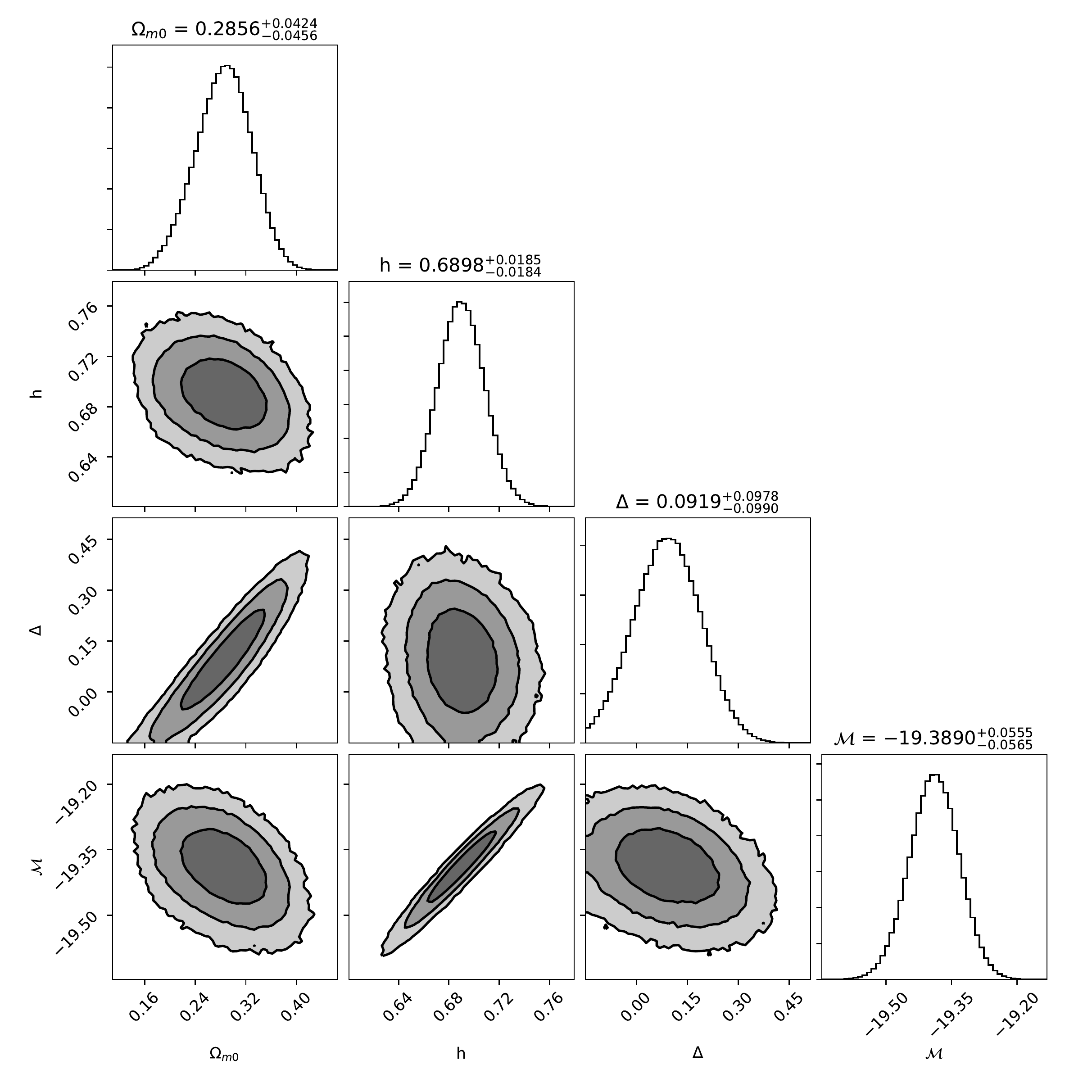}
\caption{{\it{The $1\sigma$, $2\sigma$ and $3\sigma$ likelihood contours for 
 Barrow holographic dark energy, in the case where we fix the model parameter 
${C}=3$ in $M_p$ units, using SNIa and $H(z)$ data. Additionally, we 
present the involved 
1-dimensional (1D) marginalized posterior distributions and the  parameters mean
values   corresponding to the $1\sigma$ area of the MCMC chain. 
 ${\cal{M}}$ is the usual free parameter of SNIa data that quantifies 
possible astrophysical
systematic errors,  \cite{Scolnic:2017caz}. For these 
fittings we 
obtain $\chi^2_{min}/dof = 0.8031 $.}}}
\label{Cfixed}
\end{figure}

\begin{figure}[ht]
\centering\includegraphics[width=0.98\textwidth]{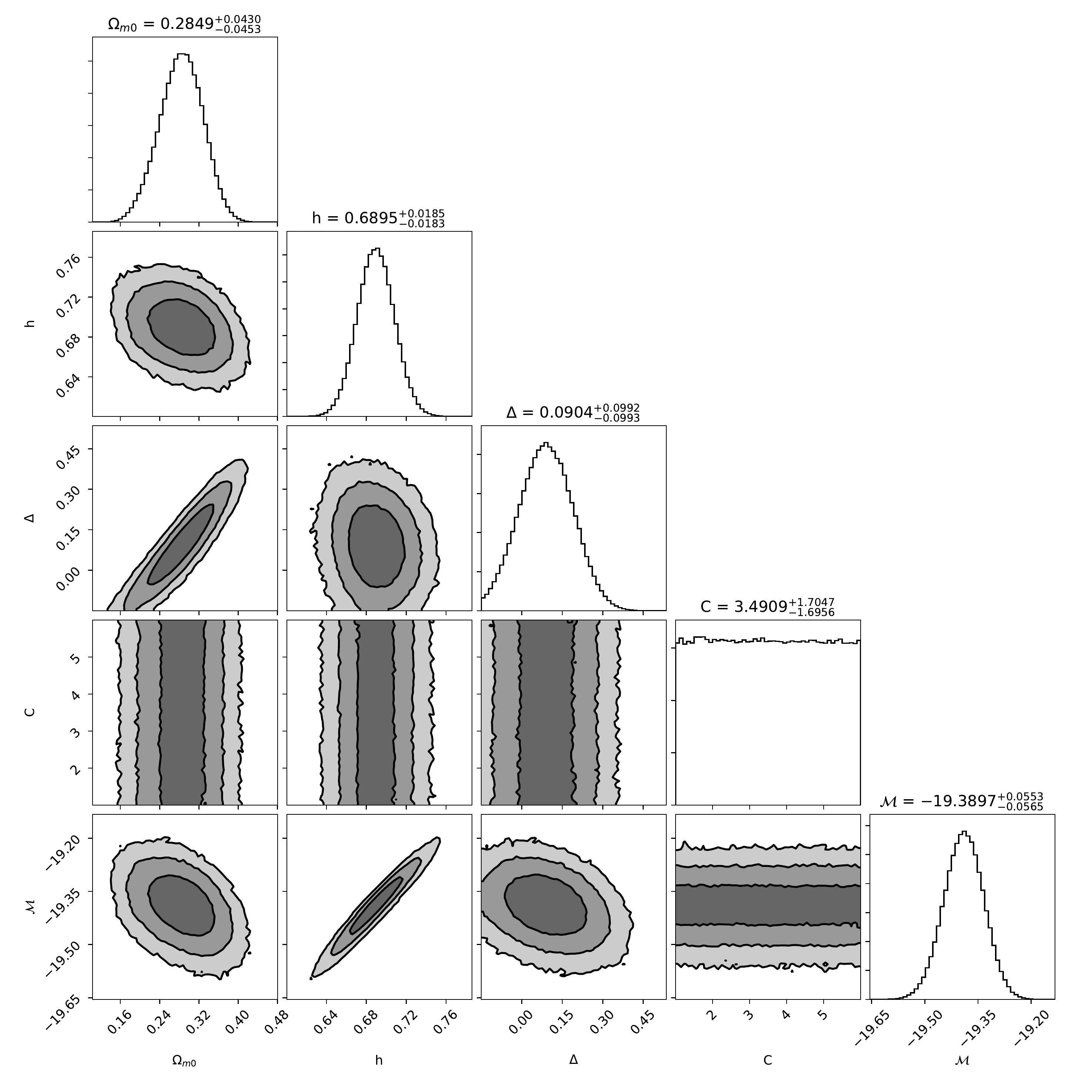}
\caption{{\it{The $1\sigma$, $2\sigma$ and $3\sigma$ likelihood contours for 
 Barrow holographic dark energy, in the case where both $\Delta$ and  $C$ are 
free 
parameters, using SNIa and $H(z)$ data. Additionally, we 
present the involved 
1-dimensional (1D) marginalized posterior distributions and the  parameters mean
values   corresponding to the $1\sigma$ area of the MCMC chain. 
 ${\cal{M}}$ is the usual free parameter of SNIa data that quantifies 
possible astrophysical
systematic errors  \cite{Scolnic:2017caz}. For these 
fittings we 
obtain $\chi^2_{min}/dof =  0.8179$.}}}
\label{Cfree}
\end{figure}

In  Table \ref{tab:Results1} we summarize the results for 
the parameters. Moreover, in  Figs. \ref{Cfixed} and 
 \ref{Cfree} we present the corresponding  likelihood contours. 
In the case where $C$ is kept fixed, we observe that 
$\Delta=0.095_{-0.100}^{+0.093} $. 
As we can 
see, the standard value $\Delta=0$ is inside the 1$\sigma$ region, however the 
mean value is $\Delta=0.095$ and thus a deviation from the standard case is 
preferred. Furthermore, we can see that $h=0.6895_{-0.0189}^{+0.0187} $ i.e we 
obtain an
$H_0$ value close to
  the Planck one $H_0 = 67.37 \pm 0.54 \ \mathrm{km \, s^{-1} \, 
Mpc^{-1}}$ \cite{Aghanim:2018eyx} instead to the direct value
$H_0 = 74.03 \pm 1.42 \ 
\mathrm{km \, s^{-1} \, Mpc^{-1}}$ \cite{Riess:2019cxk}, which was somehow 
expected since   
    the Hubble parameter is constrained only from the CC data, since the 
distance modulus from supernovae Ia cannot directly constrain $H_0$.

In the case where both $\Delta$ and  $C$ 
are free parameters, we observe that $\Delta=0.094_{-0.101}^{+0.093}  $, which 
is quite similar with the previous $C$-fixed case.   This implies that 
the deformation exponent $\Delta$ is constrained not to have its standard 
value, i.e. deviation from standard holographic dark energy is slightly 
favored. Concerning the parameter $C$ we 
find that $ 3.423_{-1.611}^{+1.753}$.
Finally, for the Hubble rate we obtain $ h = 0.6892_{-0.0189}^{+0.0187}$
and thus, similarly to the fixed-$C$ case, it is close to the Planck value.

As a final step, we test the statistical significance of the above 
constraints, 
implementing the AIC, BIC and DIC criteria described above. In  particular, we 
compare the two versions of Barrow holographic dark energy, namely the one with 
$C$ fixed and the one with both $\Delta$ and $C$ left as free parameters, with
  the concordance $\Lambda$CDM paradigm, and in
 Table \ref{tab:criteria} we depict the results. 
 As we observe, $C$-fixed Barrow holographic dark energy is more efficient than 
 the $C$-free scenario, as the extra free parameter does not contribute in the 
fit. This becomes evident 
 from Fig. \ref{Cfree}, where the 1$\sigma$ area of the parameter $C$ is not 
closed. Due to the latter fact, the DIC criterion cannot quantify well the 
adequacy of the $C$-free model. Thus, it is imperative to use AIC to proceed 
with model selection. However, to compare the other two models, one can still 
  use DIC.  As $\Delta DIC$ is smaller than 2, $C$-fixed and  $\Lambda$CDM 
are statistically equivalent. Using AIC to compare all models used here, we 
find that $C$-free model is in middle tension with $\Lambda$CDM while $C$-fixed 
is statistically equivalent with $\Lambda$CDM.
  Finally, $\Lambda$CDM paradigm seems to be slightly 
   statistically preferred.

\begin{table}[ht]
\tabcolsep 4.0pt
\vspace{1mm}
\begin{center}
 \begin{tabular}{ccccccc} \hline \hline
Model & AIC & $\Delta$AIC & BIC &$\Delta$BIC & DIC & $\Delta$DIC
 \vspace{0.05cm}\\ \hline
\hline

BHDE$|_{_{C fixed}}$ &  62.449  & 2.088 & 70.894 & 4.103 & 61.591 & 1.683 \\  
BHDE$|_{_{C free}}$ &   64.901   & 4.540 &   75.292 & 8.501 & 61.118 & 1.210 \\ 
$\Lambda$CDM & 60.361 & 0 & 66.791 & 0 & 59.908 & 0  \\ 
\hline\hline
\end{tabular}
\caption{The information criteria 
AIC, BIC and DIC for the examined cosmological models,
along with the corresponding differences
$\Delta\text{IC} \equiv \text{IC} - \text{IC}_{\text{min}}$.
\label{tab:criteria}}
\end{center}
\end{table}




\section{Conclusions}
\label{Conclusions}

 In this work used observational data from Supernovae (SNIa) Pantheon sample, 
as 
well
 as from direct measurements of the Hubble parameter from the cosmic 
chronometers (CC) sample,
 in order to extract constraints on the scenario of Barrow holographic dark 
energy.
 The latter is a new holographic dark energy scenario which is based
 on the recently proposed Barrow entropy, which arises from the modification of 
the 
black-hole surface due to quantum-gravitational effects. In particular, the 
deformation from standard
Bekenstein-Hawking entropy is quantified by the new exponent $\Delta$,
with $\Delta=0$ corresponding to standard case, while $\Delta=1$ to maximal 
deformation.
Hence, for $\Delta=0$ Barrow holographic dark energy coincides with standard 
holographic dark energy, 
while for $0<\Delta<1$ it corresponds to a new cosmological scenario that 
proves 
to 
lead to interesting and rich behavior \cite{Saridakis:2020zol}. 
Lastly, in the limiting case $\Delta=1$    one obtains 
$\rho_{DE}=const.=\Lambda$ and hence $\Lambda$CDM 
paradigm is restored, through a a completely different physical framework.
  
We first considered the case where the new exponent $\Delta$ is the sole 
model parameter, in order to investigate its pure effects, i.e. we
fixed the model parameter $C$ to 
its value for which  Barrow 
holographic dark energy restores exactly   standard holographic dark energy 
in the limit $\Delta=0$. As we showed, the standard value $\Delta=0$ 
is inside the 1$\sigma$ region, however the 
mean value is $\Delta=0.094$, namely a deviation is favored. Additionally,
for the Hubble rate we obtained a value  $ 0.6895_{-0.0189}^{+0.0187}  $   
close 
to
  the Planck    instead to the direct value, which was   expected since   
    the Hubble parameter is constrained only from the CC data, since the 
distance modulus from supernovae Ia cannot directly constrain $H_0$.

In the case where we let both $\Delta$ and $C$ to be free model parameters,
we found that   $ 0.094_{-0.101}^{+0.094} $ , and hence  
deviation 
from  standard holographic dark energy is preferred.
  Concerning the Hubble rate we found that it is close to the Planck value too.

Finally, we performed a comparison of Barrow holographic dark energy with 
 the concordance $\Lambda$CDM paradigm, using the 
  AIC, BIC and DIC information criteria. As we showed, the one-parameter scenario
  is statistically compatible with $\Lambda$CDM, and preferred
comparing to the two-parameter one.
In summary,  Barrow holographic dark energy is in agreement with
cosmological data, and it can serve as a good candidate for the description 
of nature.

\providecommand{\href}[2]{#2}\begingroup\raggedright\endgroup
\end{document}